# Walking Through the Method Zoo: Does Higher Education really meet Software Industry Demands?

Marco Kuhrmann[1], Joyce Nakatumba-Nabende[2], Rolf-Helge Pfeiffer[3],
Paolo Tell[4], Jil Klünder[5], Tayana Conte[6], Stephen G. MacDonell[7] and Regina Hebig[8]

[1]Clausthal University of Technology, Goslar, Germany, Email: marco.kuhrmann@tu-clausthal.de
[2]Makerere University, Kampala, Uganda, Email: jnakatumba@cis.mak.ac.ug
[3,4]IT University of Copenhagen, Copenhagen, Denmark, Email: {ropf, pate}@itu.dk
[5]Leibniz University Hannover, Hannover, Germany, Email: jil.kluender@inf.uni-hannover.de
[6]Federal University of Amazonas, Manaus, Brazil, Email: tayana@icomp.ufam.edu.br
[7]Auckland University of Technology, Auckland, New Zealand, Email: stephen.macdonell@aut.ac.nz
[8]Chalmers | University of Gothenburg, Gothenburg, Sweden, Email: regina.hebig@cse.gu.se

**Abstract**

*Software engineering educators are continually challenged by rapidly evolving concepts, technologies, and industry demands. Due to the omnipresence of software in a digitalized society, higher education institutions (HEIs) have to educate the students such that they learn how to learn, and that they are equipped with a profound basic knowledge and with latest knowledge about modern software and system development. Since industry demands change constantly, HEIs are challenged in meeting such current and future demands in a timely manner. This paper analyzes the current state of practice in software engineering education. Specifically, we want to compare con- temporary education with industrial practice to understand if frameworks, methods and practices for software and system development taught at HEIs reflect industrial practice. For this, we conducted an online survey and collected information about 67 software engineering courses. Our findings show that development approaches taught at HEIs quite closely reflect industrial practice. We also found that the choice of what process to teach is sometimes driven by the wish to make a course successful. Especially when this happens for project courses, it could be beneficial to put more emphasis on building learning sequences with other courses.*

## 1. INTRODUCTION

Given the huge and growing number of concepts, methods, and technologies available for software and system development, many educators have to make decisions over what to include in or exclude from a curriculum and what not. In software engineering, among other things, the selection of software development frameworks, methods and practices constitutes a major challenge. Is it better to focus on the traditional models to thoroughly teach software engineering foundations? Is it better to teach "just enough" of methodology to serve a specific course or project context? Or is it better to largely ignore the fashions and teach the students how to learn and adapt any approach?

Software and system development is thus inherently diverse and a still growing number of application domains vitally depend on software. Consequently, software and system development has to fulfill a number of context-driven requirements.

For instance, in the field of dependable systems, software is critical and has to comply with standards. Cars or medical devices are developed following such standards [1], [2], which "dictate" the way software is developed to a large extent, e.g., by defining quality management systems or test-, integration-, evaluation- and approval procedures. At the other end of the spectrum, startups work in the most pragmatic way possible to try out new ideas and to release a product as quickly as possible. Wasserman [3] states that *"...many startups [...] are completely unsystematic developing their MVP [minimum viable product], following coding practices most accurately described as 'hacking.' Products developed this way are often poorly architected, lack documentation..."*. He continues that, as soon as these startups attracted funding and customers, they need to grow and, hence, to establish all those procedures they skipped before to allow for a sustainable business. These extremes illustrate the field of tension in which educators must teach students: an extensive elaboration on standards for critical systems seems to be too much, but addressing the "hacking"-only dimension seems to be too little. Recent research confirms that companies use a multitude of different development approaches, which they combine in so-called *hybrid methods* [4], [5]. That is, in higher education, students should be prepared for this very situation, i.e., different development approaches should be subject to teaching.

***Problem Statement and Objective:*** Given that it typically takes three to five years to educate students, it is hard for HEIs to catch up with industrial innovation cycles and, moreover, it is impossible to teach everything in just one course. Therefore, we analyze the current state of practice in software engineering education and compare it with industrial practice. Specifically, we consider whether software development frameworks, methods and practices as used in industry are present in HEI courses. Our study aims to draw a big picture providing support for educators to evolve their courses to strategically balance basic knowledge and industry demands.

***Contribution:*** Based on an international online survey through which we collected information about 67 software engineering courses, we studied the use of different frameworks, methods and practices for software and system development. We analyze the data in the context of recent industry-related research on the use of hybrid methods for software and system development [5], [6] to draw a big picture and to evaluate the coverage of today's teaching in the context of industrial practice. Our analysis shows that development approaches taught in HEIs reflect industrial practice reasonably well. We also found that the choice of what process to teach is sometimes driven by the wish to make a course successful. Especially when this happens for project courses, it could be beneficial to put more emphasis on building learning sequences with other courses.

***Outline***: The rest of the paper is structured as follows: Section 2 presents related work. Section 3 describes the research design and Section 4 presents the results. A discussion of the results together with the threats to validity is presented in Section 5, before we conclude the paper in Section 6.

## 2. RELATED WORK

In 2000, Shaw [7] observed changes in software development, revisited the current state of software engineering education, and identified four main challenges educators face: (i) identifying distinct roles in software development and providing appropriate education for each, (ii) instilling an engineering attitude in educational programs, (iii) keeping education current in the face of rapid change, and (iv) establishing credentials that accurately represent ability. She proposed to provide education for each of the different roles in the software development process. It is insufficient, if a student only knows how to write code. Other activities in the software lifecycle, such as requirements engineering, are also important. She suggested to teach from an engineering point of view and proposed several aspects to improve software engineering courses and to meet the required changes in the development process, e.g., the consideration of good examples and the involvement of end users. To address the different requirements of software engineering curricula, a number of curriculum guidelines have been published [8]–[10]. These guidelines either address software engineering in its entirety, e.g., [8], [9], or focus on specific aspects such as global software engineering [10].

In the context of *active learning* principles (e.g. Dale's Cone of Learning [11]), a close collaboration with industry in joint projects becomes more and more popular. Brügge et al. [12] provide insights into years of collaborative projects in which students get deep insights into technology, product development, and team work. Mahnic [13] presents lessons learned and recommendations for integrating agile software development in a capstone course. The literature about integrating practical elements in software engineering courses is rich, e.g., [14]–[18], and covers a multitude of development approaches and techniques, such as agile software development techniques [19], [20] or software quality and testing [21]. Besides the development approaches used to run projects, specific methods are also studied for their suitability to support learning. For instance, Mendes et al. [19] study pair programming as an approach to teach students programming. Based on 300 computer science students, they found pair programming an effective approach to teach programming and also for software design. Hence, from the educator's perspective, the decision for or against a specific development approach or technique is driven by the practical relevance *and* the suitability for achieving the learning goals.

The relevance of a single development approach is, however, hard to determine in an absolute sense. Over the years, two main streams have been distilled: the *traditional* processes and the *agile* methods. In 2011, West et al. [22] claimed that these two worlds are not separated, but rather integrated by companies in *hybrid methods* for which West et al. coined the term "Water-Scrum-Fall". Companies use a variety of different software and system development frameworks, methods and practices in combination as shown by [4], [23]. To the best of our knowledge, there is no study available, which analyzes the state of practice comparing the development approaches taught in software engineering education similarly to their use in industry. This paper thus fills a gap in literature by providing a study on the use of the different frameworks, methods and practices for software system development and an analysis of the methodological coverage of industrial practice.

## 3. RESEARCH METHOD

The overall research was organized according to the schema shown in Fig. 1. It is based on a comparison of a survey of HEI educators and industry-research project data. The industry-related data was provided by the second stage of the HELENA study [5], [6].

In the following, we describe the method in detail by presenting our research questions in Section 3-A, our instrument and data collection procedures in Section 3-B, and our analysis procedures in Section 3-C.

### A. Research Objective and Research Questions
Our overall research objective is to analyze the current state of practice in software engineering education and to

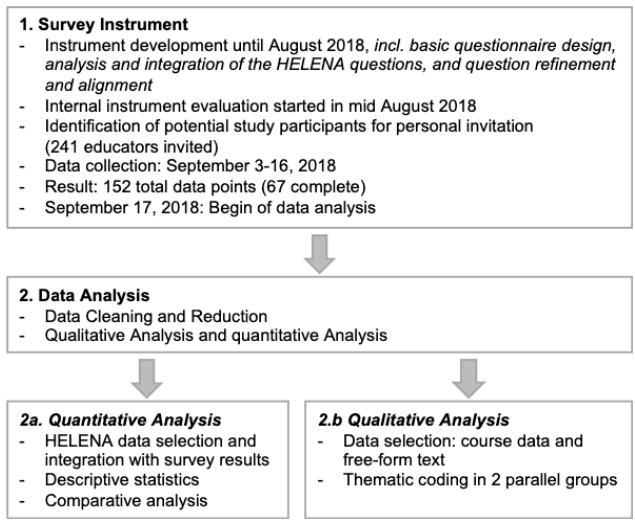

Fig. 1. Overview of the research method implemented in this study.

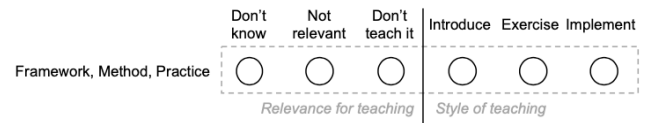

Fig. 2. Categories used to evaluate the different approaches to software and system development for educational purposes.

compare it with industrial practice to build an understanding if frameworks, methods and practices for software and system development taught at universities reflect industrial practice. To address our objective, we study the following research questions:

**RQ1:** *Which software and system development frameworks, methods and practices are taught in higher education?* This research question aims at determining the current state of practice in teaching the different software and system development approaches.

**RQ2:** *To what extent does higher education cover and coincide with software and system development frameworks, methods and practices that are applied in industry?* In this research question, we study if and how closely the currently taught software and system development approaches relate to industrial practice.

**RQ3:** *Why do educators decide to cover the chosen development frameworks, methods and practices?* In this research question, we study reasons why educators decide to (not) teach specific software and system development frameworks, methods and practices.

**B. Instrument Development and Data Collection**

To collect data, we use the *survey instrument* [24]. We designed an online questionnaire to collect data from educators in higher education about the software and system development frameworks, methods and practices they use in their courses. Hence, the *unit of analysis* was a course related to software and system development.

*1) Instrument Development:* In order to know which software and system development frameworks, methods and practices that are applied in industry, we accessed the data from the HELENA (Hybrid dEveLopmENt Approaches in software systems development) project. HELENA is a large-scale multi-staged international survey project involving more than 70 researchers from more than 25 countries [4]–[6]. The HE- LENA study yielded 1,467 data points reporting the current state of practice in modern software systems development and confirms the claim by West et al. [22] that hybrid development methods have become the norm in software systems development.

To obtain comparable insights, we developed the educator questionnaire based on HELENA's publicly available code- book [6]. Specifically, we adopted the two questions PU09 (*Which of the following frameworks and methods do you use?*; 24 items plus an optional text question to add further methods) and PU10 (*Which of the following practices do you use?*; 36 items plus an optional text question to add further practices), and developed a new set of answer options based on categories from Dale's Cone of Learning [11] to map to the original categories from the HELENA study (Fig. 3). The categories shown in Fig. 2 cover two components: (i) the relevance for teaching and (ii) the teaching style. The three *teaching style* categories are defined as follows:

**Introduce** This category covers frameworks, methods or practices that are introduced only, e.g., by naming them on lecture slides and providing a brief overview.

**Exercise** This category covers teaching of the framework, method or practice including discussions, talks about that framework, method or practice by students, or its use in (practical) exercises.

**Implement** This category includes the practical implementation of a framework, method or practice, e.g., in simulations and collaboration and/or team-based projects, such that students practically apply it.

A team of four researchers used the questions taken from the HELENA questionnaire to develop the rest of the educator questionnaire. The other four researchers involved were tasked to thoroughly review the questionnaire. In August 2018, the questionnaire was internally released for technical evaluation (Fig. 1). The questionnaire's language was English.

*2) Instrument Structure:* The final questionnaire consists of four parts: (i) Demographics (five questions), (ii) Course Information (eight questions), (iii) Process Use (seven questions), and (iv) Closing (five questions). In total, the questionnaire consisted of a maximum of 25 questions (most questions were optional and/or only shown depending on previous answers). Detailed information can be taken from [25].

*3) Data Collection:* The data collection period was September 3–16, 2018. Participants in the study were selected twofold: a first set of participants was created from the involved researchers' networks. A second set was formed from the authors of the 2017 and 2018 editions of the *ICSE Software Engineering Education and Training Track*. After removing duplicate contacts, finally, 241 individuals were contacted through personal invitation e-mails. Each participant was provided with a personal code. If a participant wanted to share experiences on more than one course, this code was used to link all courses to each other. From the 241 educators contacted, 152 started and 63 executed the questionnaire completely (response rate:

26.1%) and provided 67 cases. The full (anonymized) data can be taken from [25].

**C. Data Analysis**

As shown in Fig. 1, the first step was the data reduction and cleaning. Specifically, we analyzed the raw data for tool-specific *NA* and *-9* values. *NA* values indicate that participants did not answer an optional question, while *-9* values indicate missing information due to a skipped question. Further, we identified those participants that provided multiple cases and ensured all (meta-)data being consistent. The cleaned dataset was used for the quantitative and the qualitative analysis.

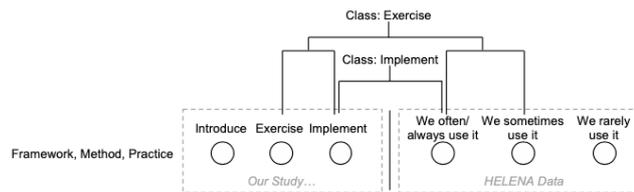

**Fig. 3.** Integration of the two data sources into the classes "Implement" and "Exercise" for the statistical analysis.

*Quantitative Analysis*: The first step in the quantitative analysis was to run the descriptive statistics on the dataset. In parallel, the provided HELENA raw dataset was analyzed and the data related to our research was extracted and prepared for further use. For the general analysis of the use of the different frameworks, methods and practices, we used simple tables containing absolute and relative numbers (Table II and III). We computed a usage ranking based on our survey data.

Beyond the plain numbers, we are especially interested if the most frequently used frameworks, methods and practices are also those most intensively taught and exercised ones in teaching. For this, we extracted the data from the two data sources and built two classes as illustrated in Fig. 3. The first class *Implement* contains all frameworks, methods and practices, i.e., a list of 60 different development approaches in total, that are often/always used in practice and that the educators stated were being taught in the *implement*-style (Fig. 2). The purpose of this class is to study if those frameworks, methods and practices most frequently used in practice are also most intensively taught. In the second class *Exercise*, we also include the "less frequently" used frameworks, methods and practices, i.e., we study if those development approaches that are (at least) sometimes used are (at least) taught in the *exercise*-style. We chose the non-parametric Kendall's τ to test for association of the data. The quantitative analysis was designed by three researchers of which, finally, two executed the analysis and the third researcher checked the procedures and results.

*Qualitative Analysis:* A qualitative analysis was performed for two aspects of the study. The first aspect was concerned with the characterization of the courses that were reported by the participants. The second aspect was concerned with studying the reasons why educators decided for or against a specific framework, method or practice. The characterization of the courses (13 themes derived from course descriptions and keywords) was conducted by one researcher and reviewed/refined by a second one. The analysis of the reasons for (not) teaching frameworks, methods and practices (10 themes derived from the free-text answers) was executed by one researcher and reviewed by a second one. Both qualitative analyses were quantified and presented as charts and tables. The thematic analysis of the reasons for (not) teaching frameworks, methods and practices was complemented with selected statements provided by the participants.

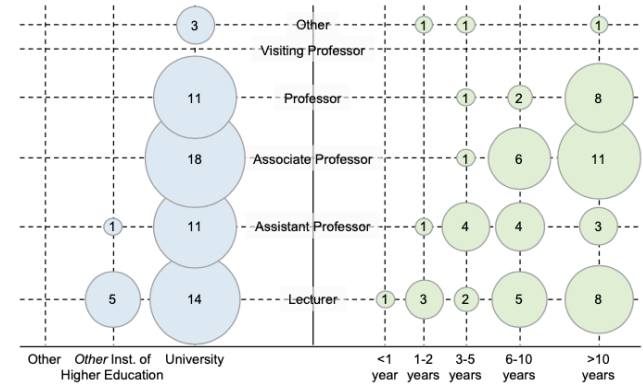

**Fig. 4.** Characterization of the 63 participants of the survey.

TABLE I. THEMES EXTRACTED FROM PARTICIPANT-PROVIDED KEYWORDS AND COURSE DESCRIPTIONS (n = 64).

| Code | Content | n |
| --- | --- | --- |
| DEV | General software development, incl. Java etc. | 24 |
| SE | Software Engineering mostly introduction courses | 22 |
| PROJECT | Project courses, e.g., industry projects and cases | 21 |
| QUALITY | Quality management and software test, e.g., TDD | 21 |
| ARCH | Architecture, patterns, agents, etc. | 18 |
| PROC | Software processes, incl. maturity models | 16 |
| AGILE | Agile and lean software development | 13 |
| MODELLING | Modelling, incl. UML, MDA etc. | 11 |
| TEAM | Team work in general | 10 |
| POM | Project management (general) | 8 |
| RE | Requirements Engineering activities | 8 |
| PRODUCT | Product development, incl. IoT, product lines etc. | 5 |
| FORMAL | Formal methods, algorithms, data structures etc. | 2 |

## 4. RESULTS

After characterizing the study population, we present the results organized according to the research questions.

*Study Population:* In total, 63 participants provided 67 data points. Figure 4 gives an overview of the participants of which 57 are employed at a university (90.48%) and six are employed at other institutions of higher education (9.52%). The participants can be considered experienced teachers: 31 (49.21%) have more than 10 years of teaching experience, 17 (26.98%) have between six and 10 years, and the remaining approx. 24% have less than five years of experience. Educators from 17 countries provided data: 20 from Denmark, 10 from Brazil, nine from Germany, four from Sweden, three from Chile, Italy and the United States, two from New Zealand, and one answer each was received from Ethiopia, Canada, Finland, India, Israel, Rwanda, Turkey, Uganda, and the United Kingdom.

*Program and Course Overview:* We received information about 67 courses of which 42 are mandatory in their respective study programs, the rest are semi-elective or elective. A majority of 52 courses targets Bachelor students whereas 15 courses are also open for other levels, i.e.,

TABLE II. ORDERED LIST OF FRAMEWORKS/METHODS COLLECTED IN THIS STUDY (N=67). FOR EACH FRAMEWORK/METHOD, THE USAGE DATA FROM THE HELENA STUDY (N=732) IS ALSO PROVIDED.

| Rank | Framework/Method | This study n | This study % | HELENA n | HELENA % |
|---|---|---|---|---|---|
| 1 | Iterative Development | 60 | 89.55 | 557 | 76.09 |
| 2 | Classic Waterfall Process | 44 | 65.67 | 404 | 55.19 |
| 3 | Scrum | 43 | 64.18 | 597 | 81.56 |
| 4 | Extreme Programming (XP) | 40 | 59.70 | 368 | 50.27 |
| 5 | Kanban | 28 | 41.79 | 468 | 63.93 |
| 6 | Spiral Model | 28 | 41.79 | 143 | 19.54 |
| 7 | V-shaped Process | 27 | 40.30 | 191 | 26.09 |
| 8 | Rational Unified Process | 25 | 37.31 | 143 | 19.54 |
| 9 | Domain-driven Design | 24 | 35.82 | 234 | 31.97 |
| 10 | Feature-driven Development | 23 | 34.33 | 263 | 35.93 |
| 11 | DevOps | 22 | 32.84 | 398 | 54.37 |
| 12 | Lean Software Development | 19 | 28.36 | 290 | 39.62 |
| 13 | Model-driven Architecture | 19 | 28.36 | 185 | 25.27 |
| 14 | Team Software Process (TSP) | 12 | 17.91 | 103 | 14.07 |
| 15 | ScrumBan | 10 | 14.93 | 206 | 28.14 |
| 16 | Large-scale Scrum (LeSS) | 8 | 11.94 | 157 | 21.45 |
| 17 | Phase/Stage-gate Model | 8 | 11.94 | 128 | 17.49 |
| 18 | Scaled Agile Framework (SAFe) | 8 | 11.94 | 145 | 19.81 |
| 19 | Personal Software Process (PSP) | 5 | 7.46 | 92 | 12.57 |
| 20 | Dynamic Systems Development Method (DSDM) | 4 | 5.97 | 58 | 7.92 |
| 21 | Structured Systems Analysis and Design Method (SSADM) | 4 | 5.97 | 94 | 12.84 |
| 22 | PRINCE2 | 3 | 4.48 | 65 | 8.88 |
| 23 | Crystal Family | 1 | 1.49 | 32 | 4.37 |
| 24 | Nexus | 1 | 1.49 | 64 | 8.74 |

TABLE III. ORDERED LIST OF PRACTICES COLLECTED IN THIS STUDY (N=67). FOR EACH PRACTICE, THE USAGE DATA FROM THE HELENA STUDY (N=732) IS ALSO PROVIDED.

| Rank | Practices | This study n | This study % | HELENA n | HELENA % |
|---|---|---|---|---|---|
| 1 | User Stories | 54 | 80.60 | 579 | 79.10 |
| 2 | Architecture Specifications | 49 | 73.13 | 564 | 77.05 |
| 3 | Prototyping | 47 | 70.15 | 625 | 85.38 |
| 4 | Coding Standards | 46 | 68.66 | 666 | 90.98 |
| 5 | Refactoring | 46 | 68.66 | 608 | 83.06 |
| 6 | Code Review | 44 | 65.67 | 681 | 93.03 |
| 7 | Pair Programming | 44 | 65.67 | 473 | 64.62 |
| 8 | Use Case Modeling | 44 | 65.67 | 436 | 59.56 |
| 9 | Test-driven Development | 42 | 62.69 | 453 | 61.89 |
| 10 | Continuous Integration | 41 | 61.19 | 593 | 81.01 |
| 11 | Iteration/Sprint Reviews | 41 | 61.19 | 593 | 81.01 |
| 12 | Iteration Planning | 40 | 59.70 | 579 | 79.10 |
| 13 | Automated Unit Testing | 39 | 58.21 | 616 | 84.15 |
| 14 | Backlog Management | 37 | 55.22 | 592 | 80.87 |
| 15 | Continuous Development | 36 | 53.73 | 521 | 71.17 |
| 16 | Collective Code Ownership | 33 | 49.25 | 436 | 59.56 |
| 17 | Detailed Designs/Specifications | 33 | 49.25 | 510 | 69.67 |
| 18 | Retrospectives | 33 | 49.25 | 549 | 75.00 |
| 19 | Burn-down Charts | 32 | 47.76 | 488 | 66.67 |
| 20 | Daily Standup | 32 | 47.76 | 559 | 76.37 |
| 21 | Definition of Done/Ready | 32 | 47.76 | 502 | 68.58 |
| 22 | Design Reviews | 32 | 47.76 | 581 | 79.37 |
| 23 | Release Planning | 32 | 47.76 | 633 | 86.48 |
| 24 | Expert Estimation | 30 | 44.78 | 505 | 68.99 |
| 25 | End-to-End Testing | 29 | 43.28 | 582 | 79.51 |
| 26 | Automated Code Generation | 26 | 38.81 | 461 | 62.98 |
| 27 | On-Site Customer | 25 | 37.31 | 348 | 47.54 |
| 28 | Limit Work-in-Progress | 23 | 34.33 | 440 | 60.11 |
| 29 | Velocity-pased Planning | 17 | 25.37 | 327 | 44.67 |
| 30 | Model Checking | 15 | 22.39 | 176 | 24.04 |
| 31 | Scrum-of-Scrums | 15 | 22.39 | 290 | 39.62 |
| 32 | Formal Specification | 14 | 20.90 | 348 | 47.54 |
| 33 | Formal Estimation | 13 | 19.40 | 505 | 68.99 |
| 34 | Security Testing | 13 | 19.40 | 495 | 67.63 |
| 35 | Destructive Testing | 4 | 5.97 | 285 | 38.93 |
| 36 | Automated Theorem Proving | 3 | 4.48 | 101 | 13.80 |

Master, post-graduate and "other" students. The class sizes range from eight to approx. 600 students (we had to exclude one data point due to invalid values) with a median of 40 participants. For 64 courses, participants provided information about the course content (via short text or an online link to the course description, and complemented by a keyword list). We developed 13 codes and made 179 assignments to the courses for describing the respective focal points (Table I). It has to be noted that notably courses in the broad category *SE* often cover different software engineering activities and processes. Hence, if Table I for instance names eight courses in the category *RE*, then this denotes courses that either exclusively deal with requirements engineering or at least put special emphasis on this discipline.

### A. RQ1: Frameworks, Methods and Practices in Education

The first research question was concerned with the current state of practice in teaching. For this, we used the list of frameworks, methods and practices from the HELENA survey as a template. We provided the educators with 24 software and system development frameworks and methods, and 36 practices (both in alphabetical order).

Table II shows the list of software and system development frameworks and methods ordered by the number of educators teaching each one. In this regard, "teach" refers to the category group *style of teaching* (Fig. 2), i.e, that a framework/method is taught at all. The table shows a separator at the 50%-level, i.e., which frameworks and methods are taught by at least 50% of the educators. In the same way, Table III presents the ranked list of taught practices.

Table II shows only four frameworks and methods taught by at least 50% of the participants. Among these frameworks and methods are the generic *Iterative Development* and the *Classic Waterfall Process*, and the two representatives of the agile methods *Scrum* and *XP*. Still well-represented are the *V-shaped processes* and the *Rational Unified Process*. These approaches are complemented with lean and agile frameworks and methods, yet, the table shows that notably scaling approaches for agile methods, e.g., *LeSS*, *SAFe* and *Nexus*, are not frequently mentioned. Regarding the practices taught, Table III provides a more diverse picture. At least 50% of the participants teach one out of 15 practices of which most are concerned with implementation-related tasks (see also the courses' characterization in Table I). However, *User Stories* and *Architecture Specifications* are the most frequently mentioned practices also showing the importance of analysis and design activities in the software lifecycle. Among the least mentioned practices are representatives of the "formal methods" family and, surprisingly, *Security Testing* (see also Figures 5 and 6).

### B. RQ2: Relation of Education and Industrial Practice

The second research question is concerned with the topics taught and the topics' relation to practically used frameworks, methods and practices in industry. To draw a big picture, we compared the 67 courses—specifically the use of frameworks, methods and practices in these courses—with the use of hybrid methods as reported in the 732 cases of the HELENA study.

*1) Visual Inspection and Comparison:* Figure 5 and Fig. 6 provide a comparison based on the use of the different development approaches. Both figures were developed based on the usage information from the category group *style of teaching* (see also Fig. 2, Table II and Table III).

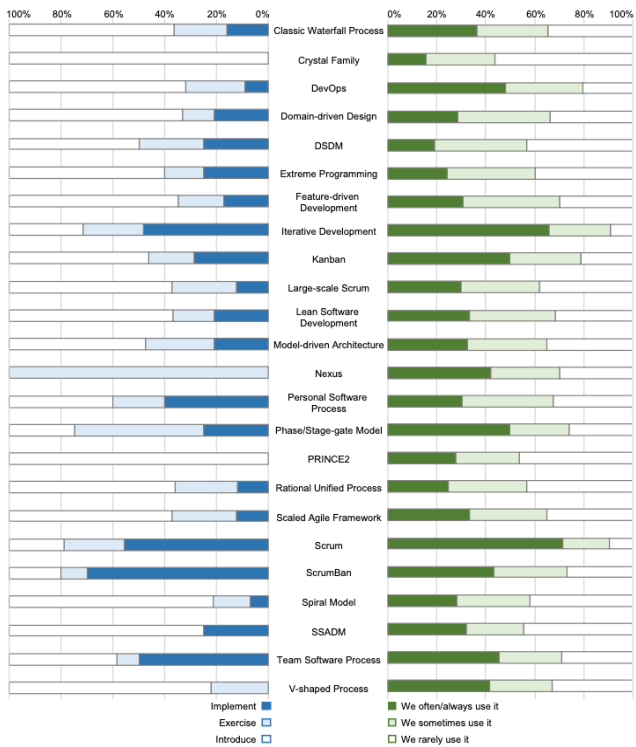

**Fig. 5.** Comparison of frameworks/methods taught (n = 67; left) and the usage style in practice (HELENA data set; n = 732; right).

The HELENA counterpart was developed from the two questions PU09 and PU10 of the HELENA data (Section 3-B1). Figure 5 and Fig. 6 provide the data normalized in an interval [0,1] expressed in percentages.

Figure 5 shows the comparison of the framework/method use. The figure shows that the most frequently used frameworks/methods in industry are also present in the curricula. That is, *Iterative Development* and *Scrum* are frequently used and also often taught. For other framework/methods, we see differences. For instance, while the *V-shaped Processes* are well-represented in the HELENA data set, few courses address this process family. For *PRINCE2*, *Nexus* and the *Crystal Family*, the picture is even more clear as these methods are barely present in teaching; if at all. An interesting behavior can be seen regarding the more recent lean and continuous development approaches, e.g., *DevOps*, *Kanban* or *Lean Software Development* in general, which are present but less frequently mentioned. The only exception is *ScrumBan*, which has a higher relative frequency in teaching than in the practical use.

Figure 6 compares the practices taught with the practices used in industry. The figure shows a good coverage of the industrially relevant practices in teaching. However, a number of practices are not as present as expected. For instance, *Formal Estimation*, *Formal Specification*, *Security Testing*, and *Velocity-based Planning* are relevant to practice, but not taught in a comparable share. Due to the limited dataset of our study, we did not have any mentions of *Automated Theorem Proving* in teaching even though, however, it is used in practice. Hence, Fig. 5 and Fig. 6 show the overall coverage of the frameworks, methods and practices taught, but also shows where industrial practice is not reflected.

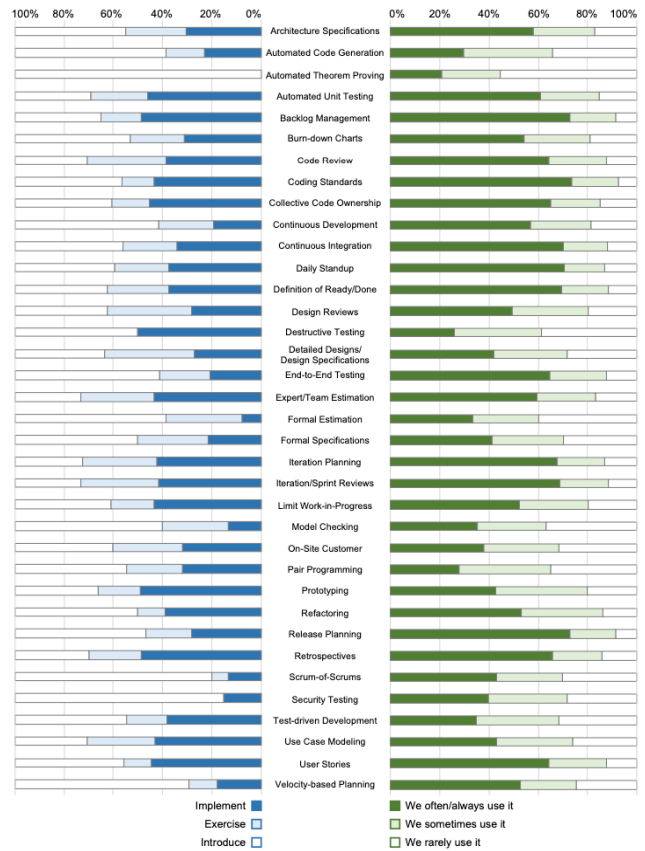

**Fig. 6.** Comparison of practices taught (n = 67; left) and the usage style in practice (HELENA data set; n = 732; right).

*2) Relation of Teaching and Practice:* To study in more detail if the topics addressed in teaching sufficiently cover the topics relevant to industry, we test the correlation of the two datasets using the non-parametric Kendall's $\tau$ test. The tests for the two classes described in Section 3-C produced the following results, which show that frameworks, methods and practices taught provide reasonably good coverage of industrial practice.

| Class | Test | $\tau$ |
|---|---|---|
| Implement | z=7.0929, p-value = $1.313 \times 10^{-12}$ | 0.6413766 |
| Exercise | z=7.3317, p-value = $2.273 \times 10^{-13}$ | 0.6581279 |

**Finding 1:** Similar to industrial practice, educators utilize a multitude of frameworks, methods and practices in teaching. Compared to industry, the bandwidth of the development approaches used is smaller and focused on few frameworks and methods and more development-oriented agile practices.

**Finding 2:** The direct comparison shows that the frameworks, methods and practices used in teaching correspond with industrial practice. Only few deviations and different focal points can be found.

**Finding 3:** A statistical analysis of our data and the HELENA data shows a strong correlation between the style of teaching and the intensity of use. That is, frameworks, methods and practices taught at universities provide a good coverage of development approaches that are used in industry.

### C. RQ3: Reasons for (not) Teaching Different Frameworks, Methods and Practices in Education

After evaluating the different software and system development frameworks, methods and practices, the participants were presented with a summary of those approaches that they stated that they did not use in teaching (Fig. 2, item "Don't teach it"). This summary was provided in the context of the question "*Why don't you teach the frameworks/methods and practices?*" Of the 63 lecturers, five did not provide an answer for their reported course, i.e., our analysis is based on 62 of the 67 courses.

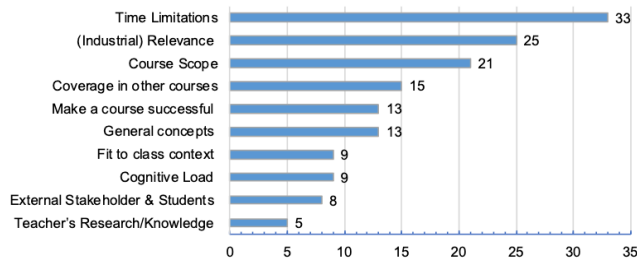

**Fig. 7.** Summary of reasons for the selection of methods, frameworks, and practices in teaching.

As described in Section 3-C, we performed a thematic coding to qualitatively analyze the reason for (not) teaching particular frameworks, methods and practices. The overall result is illustrated in Fig. 7. It is no surprise that the main reason is *time limitations*, which was mentioned for approx. half of the courses (33). However, there are further aspects driving decisions for or against the different frameworks, methods, and practices. In the following, we provide an overview of these reasons as reported by the educators, and we provide selected quotes.

*1) (Industrial) Relevance:* For 25 courses, the (industrial) relevance of a framework, method or practice was mentioned as the reason to teach or not teach it. In most cases, this relevance was linked to its use in the local industry or with regards to the dissemination and use of the process in general.

> "We base our materials on what the industry is currently using or making a shift towards and Scrum, XP, Kanban and UP are the commonly used methodologies within the industry here in Northern Denmark" [case 215]

Other educators consider development approaches relevant if they are very well-known or generally considered "essential":

> "These skills are essential to master as a modern developer, in my opinion." [case 222]

Finally, in six cases educators emphasized timeliness as a factor when selecting the frameworks, methods and practices for teaching. This can also be a reason to exclude certain methods from a course:

> "I believe some of them are historic and not used by the local industry (like Crystal, DSDM, FDD, SSADM...)." [case 202]

*2) Course Scope and Coverage in Other Courses:* Courses at HEIs must not be seen in isolation. They are usually part of comprehensive study programs. This is reflected in the observation that the courses' scope was mentioned in 21 cases as a reason to not teach a framework, methods or practice. In 14 further cases, the reason for excluding a topic was that it was covered in other courses.

*3) Teacher's Research and Knowledge:* Another factor influencing the selection of frameworks, methods and practices is the educator's own research and knowledge as mentioned for five courses. This is also related to the teaching capacities of research groups, as described by one participant:

> "The other ones are not part of the research at our research group and thus we do not focus them. We are a small research group and cannot cover all frameworks, methods and practices." [case 159]

Other educators state that they do not like to teach methods with which they have little or no experience and for which no appropriate teaching material is available:

> "In addition I have not yet found a good course book that can be effectively used together with an iterative approach. (The teaching material tends to come in a format befitting the waterfall model, discussing elicitation, design, implementation, testing and deployment in a very sequential manner." [case 204]

*4) External Stakeholders & Students:* In eight cases, external stakeholders (i.e., practitioners acting as partners in a course) and the students as such were named factors influencing the selection of frameworks, methods and practices. For instance, if practitioners are involved in a course, their influence on the selection of the development approach ranges from "just being taken into account" to "making the choice" by providing a project with a given framework:

> "...The mentors teach whatever frameworks, methods & practices they are currently using in their active open-source projects. These projects are not created specifically for teaching but are active development projects with a real community..." [case 152]

In some project courses, the choice of frameworks, methods and practices is given to the students by allowing them to select between a set of given processes or even by giving them full freedom to come up with their own processes:

> "[...] We emphasize prototyping and use cases as important techniques to engage with and get information from, stakeholders. The students are told to use whatever (generic) model they find applicable and to reflect upon the outcome. In practice, however, most end up with a kind of chaotic waterfall due to time constraints and limited experience." [case 156]

*5) Cognitive Load:* The decision for or against teaching a certain framework, method or practice is made based on pedagogical considerations. In nine courses, cognitive load played a role in the decision-making process. On the one hand, many process-related concepts were considered too advanced for the students, e.g., DevOps and V-shaped processes were considered too advanced for some basic courses:

> "I consider DevOps highly relevant but that would have been too much for a bachelor course; is covered at the master level" [case 155]

On the other hand, the participants are concerned to overwhelm students with the multitude of the available frameworks, methods and practices:

> "Avoiding overwhelming students with more than two agile methods and around 30-35 agile practices (management/development)" [case 189]

*6) General Concepts over Specific Methods & Frameworks:*

Another pedagogical consideration, which was reported for 12 courses, concerns the intention to teach general concepts rather than specific methods or frameworks. This often includes the intention to provide students with more reusable knowledge that can be applied to many cases in practice:

> "Generic process models are introduced but we do not spend time on specific instances; e.g. Scrum. Nor on the various families of frameworks. We assume that the practical and guided experience with planning and conducting a complex analysis, design and development process provide the background for later studies of specific frameworks." [case 156]

One participant stated that it is not realistic to hope for teaching students the "exact and right" process, since companies use their own variants and mixtures:

> "UP, XP , Waterfall and SCRUM is enough to learn Agile and Classic methods. I companies they make their own mix anyway." [case 183]

*7) (Mis-)Fit to Class Context:* When deciding how many and complex frameworks, methods and practices to teach in a single course or in a particular class-room setup, educators have to make pedagogical decisions. We received corresponding comments from nine cases. For example, Scrum was considered to be simple enough for a class-room simulation. Other frameworks, methods and practices are often considered too complex to fit into a few-week long semester project or into class-room exercises:

> "Burn-Down Charts is a Scrum tools to plan and visualize plans in project spanning some time. My interactive exercises are executed in a very short time frames (10-20 minutes) during a class." [case 148]

> "I'd rather teach a more iterative method, however I have found it very challenging to fit this into the very short teaching periods (7-8 weeks) at my university." [case 204]

The actual course format can limit the number of processes taught as stated by the following participant:

> "No lectures in the course, only practical work, so we need to limit it to a handful of practices that work for the course scope of the course" [case 112]

*8) Methods and Practices as a Tool to Make a Course Successful:* An interesting case is courses in which frameworks, methods and practices are taught with the intention to make teaching the course as such successful—or even possible. We observed this for 13 courses. A selection is made to ensure that project courses run well:

> "Iterative Development and Prototyping are used to improve the quality of student projects, in particular with the aim of significantly increasing their chance of successfully making a project in the context of the course." [case 113]

Educators might even change processes over the years in order to make a course run more smoothly:

> "The course was previously more waterfall-ish, with separate design, implementation and test phases, but students were really struggling with the design phase and it quite often resulted in a very abstract and kind of useless design just to have something to submit for the course but not really following it in the implementation. So, we decided to merge the design and implementation and making it more iterative while also introducing other agile/scrum practices (e.g. weekly sprint review and planning meetings). In the end, we use a kind of staged model, with two weeks for requirements and project planning, then four weeks of iterative development and a final week for delivery and acceptance testing." [case 112]

In other cases, processes are taught as students will be confronted with these in later classes or in course projects:

> "Also, things like refactoring or capturing requirements (as use cases or user stories) cannot be avoided and students will face related challenges as soon as they work on their software engineering assignment..." [case 108]

Finally, single practices, e.g., model checking or TDD, might be taught to support teaching other course contents, such as concurrency or new programming languages:

> "The student already know Java, so just teaching C# without adding extra seems pointless. Introducing and actually forcing TDD has proven to be a very interesting and good way of teaching development in a new language (IMHO)" [case 199]

## 5. DISCUSSION

So, why do educators opt for teaching certain frameworks, methods and practices? Are such decisions relating to industrial practice and do the selected development approaches provide a good coverage of industrial software development? In this section, we answer the research questions and provide a discussion of findings, takeaways, and threats to validity.

### A. Answering the Research Questions

To drive our study, we posed three research questions for which we provide the answers:

*RQ1: Which software and system development frameworks, methods and practices are taught in higher education?* The data shows educators teaching a variety of development approaches. Four frameworks and methods (Iterative Development, Waterfall, Scrum and XP) and 15 practices are favored by the educators (Table II and Table III).

*RQ2: To what extent does higher education cover and coincide with software and system development frameworks, methods and practices that are applied in industry?* A visual inspection and a Kendall's τ test show that the processes taught strongly correlate with those used in industry.

*RQ3: Why do educators decide to cover the chosen development frameworks, methods and practices?* A thematic coding revealed 10 reasons for (not) teaching certain frameworks, methods and practices. The main drivers named were time constraints, course scope (incl. alignment with other courses) and pedagogical considerations. Many educators seek industrial collaboration to improve the relevance of their courses.

### B. Discussion of the Results

Our findings reveal several aspects worth a discussion. We see a variety in the teaching styles. For instance, some educators prefer more general development approaches over specific ones to ensure that students are equipped with basic knowledge that they can use for transfer to new situations. Table II provides three examples: the *Classic Waterfall*, the *Spiral Model*, and the *V-shaped Processes*, which are more intensively taught (notably in basic software engineering courses) than used in practice. Educators seem to consider these process models providing good teaching tools to explain software engineering as a whole. When it comes to the "hands-on parts", there is—respecting the limitations of the courses—little to no difference between education and practice (Table III). We argue that this happens as educators seek effective ways to establish a practically relevant and active learning environment.

Our data also shows that software and system development frameworks, methods and practices are often taught along with project courses. The basic concepts are shifted into the background as an actual process implementation to run a project course is in the spotlight. We know from previous research that there is a risk that students will focus on the course objective, i.e., the product, by skipping all (perceived) unnecessary yet important activities [26], and create a development style that Wasserman "accurately described as *hacking*" [3]. We argue that there is an opportunity to give the software process education more room. For instance, involving process teachers in all project courses could be beneficial reflect on process implementations and their impact and train software processes. This would allow for ensuring the basic principles and good practices of software engineering being continuously and consistently taught across courses:

> "It is emphatically not enough to have students read or hear about it – they need to experience for themselves what an unstructured process does to the quality of their product and the cohesion of their team and then realize how a process can help them." [case 161]

### C. Takeaways for Teachers

From the data as well as from the authors' experience, we derive some takeaways for teachers. First, we want to highlight that the teaching practice is not that far away from practice as often perceived. Yet, as HEIs have several constraints (see above), teachers should consider the following aspects:

*1) Consider other courses in your program:* Process education is ideally not just a matter of a single course in the program. When defining the process portfolio for a specific course, teachers should first answer the following questions:

- Do other courses earlier in the program already provide basic knowledge to build upon?
- Are there synergy potentials with other courses that rely on the same frameworks, methods and practices?
- Is it possible to develop "teaching sequences", i.e., to develop a consecutive program structure, where basics taught in one course lay the foundation for other courses?

*2) Less is more:* Not every course needs to teach the full set of processes available. The HELENA study [5] provides data about the industrial practice that can be used to define a proper process portfolio. Furthermore, we suggest looking into the local industry and to answer the following questions:

- Will the students be confronted with specific processes?
- Could guest lecturers from local industry provide required process knowledge?

*3) No teaching material for hybrid processes:* Lacking teaching material was found among the reasons to not teach specific processes. Furthermore, when companies use specific (hybrid) processes that do not follow any standard, teachers need to find a way to educate students without material, teach general concepts, and prepare students how to adopt new processes. We suggest teachers answering the following questions:

- Can teaching collaborations, e.g. with other HEIs, provide the required process knowledge?
- Can guest lecturers from industry provide insights into local companies' specific processes?

*4) The "Real-World Trap":* Teaching software engineering in project courses is considered a promising route towards a practice-oriented education, notably if real clients are involved in project courses [12]. Such setups, however, can lead to a situation in which the process portfolio is defined on a per- project base depending on the actual external partner. Teachers should critically and continuously (re-)evaluate the decisions made in such setups by answering the following questions:

- Do student teams really select and follow a process?
- Do industry partners really help students implement the partner's process?

### D. Threats to Validity

We discuss construct, internal, external, and conclusion validity (according to Wohlin et al. [27]) of this study.

*1) Construct validity:* As we ground our study in the HELENA study [5], [6], we are potentially affected by threats introduced by the HELENA study. A major threat to construct validity is the risk that HELENA-participants misunderstood the questions. As described in [5], a test phase and a multi- language questionnaire were utilized to mitigate this risk.

The construct validity of this study might be threatened by our classification of the teaching intensity (Fig. 2). The HELENA study uses a measure to assess the intensity of use, which, however, is not suitable for teaching-related assessments. Therefore, we developed a schema based on educational theories (Dale's Cone of Learning [11]). The assumption is that active learning approaches lead to a more intensive learning. Another threat might be introduced due to the subject selection, i.e., hand-picked educators that report on self-selected courses. This potentially provides a limited perspective on the frameworks, methods and practices taught, since certain topics of interest are part of other courses as some of the participants also noted. Yet, data collected so far provides an indication of the development approaches taught and the way of teaching. In this regard, it must not be forgotten that we use data about the *current* teaching and industrial use of different development approaches. Changes that are instantiated to teaching today will have to consider also where software industry stands in 3-5 years, which is of course hard to predict.

*2) Internal validity:* The internal validity might be threatened by the subject selection (this risk is present for the study at hand as well as the HELENA study [5]). In this study, the recruitment was based on the authors' professional networks and, notably, included educators of which the authors assumed teaching subjects of interest. To mitigate this threat, a second set of potential candidates was developed from authors, who published in the ICSE SEET track in 2017 and 2018. It has to be noted that, due to the construction of the survey, we cannot distinguish the SEET Track authors from the other participants, which could also affect the *conclusion validity*. Still, our results provide a limited perspective only as the invited educators do not cover software engineering in its entirety. Furthermore, even though our invitation resulted in an acceptable response rate of 26.1%, it is possible that there was a systematic reason for not answering/completing the questionnaire (again for both studies: educator and HELENA). Future studies with different data collection strategies are needed to confirm our findings. Finally, to mitigate potential threats to internal validity due to personal bias in the data analysis, same as in the HELENA study, we ensured that all data analysis steps have been performed in teams and checked by other researchers not involved in the analysis steps.

*3) External validity:* Due to the small sample size, our results lack generalizability. However, as educators from 17 countries provided information and some regions delivered more than one response, e.g., Scandinavia with 24 data points and Brazil with 10, we argue that our data provides a first big picture of teaching software engineering methods worldwide. However, a larger study is required to confirm our findings. The HELENA study, which is our basis, suffers from the same risk regarding the generalizability. Yet, the results presented in [5] provide a good indication of a generalizable observation. The presented study on educators and the HELENA study require a broader investigation to confirm the found trends.

*4) Conclusion validity:* Even though we found strong correlation with a significance level of $p < 0.05$, the conclusions drawn suffer from the small sample. Hence, a larger study is needed to confirm our findings.

## 6. CONCLUSION

The paper at hand provides a study on the use of frameworks, methods and practices for software and system development in teaching and compares them and the way they are taught with industrial practice. Hence, this study complements industry-related research and provides an educational perspective on the use of different development approaches.

Our findings based on 63 educators from 17 countries that reported on 67 software engineering courses show that the different development approaches taught at HEIs correspond to those used in industrial software development. That is, HEIs reasonably prepare students for the current methods and practices. Our data also shows opportunities for improvement. In some cases, processes are selected under the pressure to make a course (often a project course) successful and run smoothly. Here, educators could benefit from a more integrated teaching approach in which the development approaches of interest are taught concisely thus allowing students to continuously train the processes across their courses and in different projects.

The paper at hand provides a first big picture. Future work includes a more detailed analysis of the data together with the industrial counterparts provided by the HELENA study. Furthermore, an in-depth comparison of teaching and practice per country would be interesting to see whether educators provide for their local job markets. Similarly, it would be interesting to make a cross-comparison to investigate whether the type of educational institute or nature of the course impact the teaching choice. Finally, follow-up studies are required to discuss detailed setups with the educators to get more insights.

### ACKNOWLEDGMENTS

We want to thank all participants of the study, who shared the information about their courses. *Tayana Conte* is supported by CNPq (311494/2017-0). *Joyce Nakatumba-Nabende* was supported by the Sida/BRIGHT project 317 under the Makerere-Swedish bilateral research programme 2015-2020.